\documentclass[twocolumn]{aastex631}
\usepackage{gensymb}
\usepackage{graphicx,xcolor}

\begin{document}

\title{Study of the Blazhko type RRc stars in the Stripe 82 region using SDSS and ZTF}

\author{Vaidehi Varma}
\email{varma@astro.ncu.edu.tw}
\affiliation{Graduate Institute of Astronomy, National Central University, No. 300, Zhongda Rd., Zhongli District, Taoyuan City 320317, Taiwan}
\author{J\'ozsef M. Benk\H{o}}
\affiliation{Konkoly Observatory, Research Center for Astronomy and Earth Sciences, HUN-REN; 
            MTA Centre of Excellence, \\ Konkoly Thege M. u. 15-17., H-1121 Budapest, Hungary} 
\author{Chow-Choong Ngeow}
\affiliation{Graduate Institute of Astronomy, National Central University, No. 300, Zhongda Rd., Zhongli District, Taoyuan City 320317, Taiwan}

\begin{abstract}
RR Lyrae stars are pulsating stars, many of which also show a long-term variation called the Blazhko effect which is a modulation of amplitude and phase of the lightcurve. In this work, we searched for the incidence rate of the Blazhko effect in the first-overtone pulsating RR Lyrae (RRc) stars of the Galactic halo. The focus was on the Stripe 82 region in the Galactic halo which was studied by Sesar et al using the Sloan Digital Sky Survey (SDSS) data. In their work, 104 RR Lyrae stars were classified as RRc type. We combined their SDSS light curves with Zwicky Transient Facility (ZTF) data, and use them to document the Blazhko properties of these RRc stars. Our analysis showed that among the 104 RRc stars, 8 were rather RRd stars, and were excluded from the study. Out of remaining 96, 34 were Blazhko type, 62 were non-Blazhko type, giving the incidence rate of 35.42\% for Blazhko RRc stars. The shortest Blazhko period found was $12.808\pm0.001$~d for SDSS 747380, while the longest was $3100\pm126$~d for SDSS 3585856. Combining SDSS and ZTF data sets increased the probability of detecting the small variations due to the Blazhko effect, and thus provided a unique opportunity to find longer Blazhko periods. We found that 85\% of RRc stars had the Blazhko period longer than 200~d. 
\end{abstract}

\section{Introduction}

RR Lyrae stars (RRLs) are low-mass pulsating giant stars that have evolved away from the main sequence, and are on the intersection of the horizontal branch and the classical instability strip of the Hertzsprung–Russell diagram. They pulsate mostly radially but on occasion, non-radial modes can be observed, which make them a good source to study the stellar evolution and stellar pulsation \citep[for a review see e.g.][]{kolenberg2012rr, refId0}. Their pulsation modes and the shape of their lightcurves (LCs) are used to classify them into three different types: RRab type pulsates in the fundamental mode and has an asymmetric sawtooth shaped LC; RRc type pulsates in the first-overtone mode and has more sinusoidal LC than RRab; and RRd type pulsates in both the modes and also has a sinusoidal LC. 

In addition to the pulsation, there is a long-term variation observed in all types of RRLs. This is called the Blazhko effect which is a quasi-periodic modulation of amplitude and phase of a LC \citep{blazko1907, shapley1916}. Blazhko period ($P_{\mathrm B}$) is longer than the main pulsation period ($P$), and can be in the range of a few to a hundred days \citep{smith2004rr}, and for some, it is even longer than 1000 days. About 50\% of RRab stars show the Blazhko effect \citep{Benkő_2014, jurcsik2017}, while the incidence rate is lower for RRc stars (4\textendash19\%, \citealt{netzel2018}), and it is even lower for RRd stars \citep{soszynski2016}.  Despite the lower Blazhko incidence rate than RRab stars, Blazhko type RRc (BRRc) stars have been analysed and published in many stellar systems e.g. Galactic Bulge \citep{moskalik2003fourier, mizerski2003, netzel2018, netzel2019census}, Large Magellanic Cloud \citep{nagy2006incidence}, and Galactic globular clusters \citep{clement2001variable, jurcsik2015overtone}. Though, no such systematic study is known for the RRc stars in the Galactic halo. Since the physical origin of the Blazhko effect is still unknown, in spite of its relatively high abundance and long history, a study of any Galactic subsystems could contribute to understanding the effect.

In the halo region of the Milky Way, an equatorial strip with declination limits of $\pm 1\degree27$, and extending from R.A. $\sim$ 20h to R.A. $\sim$ 4h is known as the Sloan Digital Sky Survey \citep[SDSS,][]{york2000sloan} Stripe 82 region. \cite{Sesar_2010} (hereafter S10) studied RRLs of the area using SDSS's multi-band and multi-epoch observations. They published only the basic pulsation properties of RRLs, with no information about the Blazhko effect. We attempt to study BRRc stars using the data of S10 and Zwicky Transient Facility \citep[ZTF,][]{bellm2019zwicky, graham2019zwicky}. ZTF is a time-domain survey of the northern sky, exploring the transients and variable stars using the Palomar 48 inch Schmidt telescope that began its operation in 2018. Combining ZTF and SDSS data sets gave us $\sim23$ year-long LCs, which were long enough to explore the long Blazhko cycles.

\section{Data and analysis}

S10 catalogued 483 RRLs in total. Among them, 104 were classified as RRc type. The details pertaining to them can be found in their paper. These LCs are available for the public, and can be accessed using astroML \citep{astroML} package gatspy \citep{gatspy18371, vanderplas2015periodograms}.
In the present study, we used the $g$-band data. The amplitudes of RR Lyrae LCs are known to be higher in $g$-band than in the other bands, and that gives us the best opportunity to detect the smallest changes produced by the Blazhko effect. 

The ZTF LCs were retrieved also in $g$-band using the data from the DR15. The $g$ filters of SDSS and ZTF are not identical but both can be calibrated to Pan-STARRS magnitude system. ZTF pipeline calibrates ZTF magnitudes to that of Pan-STARRS system \citep{masci2018zwicky}, and \cite{tonry2012pan} proved a way to calibrate SDSS magnitudes to that of Pan-STARRS. We used the linear form of their proposed formula in order to ease the error propagation:
\begin{equation}
g_{\mathrm{ps1}}-g_{\mathrm{SDSS}} = -0.012 -0.139(g-r)_{\mathrm{SDSS}} 
\end{equation}
Finally, the combined LCs had been constructed that extended over a period of about 23 years, of which 13 years were covered by observations. The number of data points varied between 139 and 1316, with an average of 376.

Period04 package \citep[P04,][]{lenz2004period04} was used to analyse the combined LCs. 
We also coded a Python script for an alternative analysis. Using the lombscargle package of SciPy \citep{2020SciPy-NMeth}, we searched for the dominating frequencies from the LCs. These frequencies were then fitted to a LC using the Fourier series:
\begin{equation}
    y = A_0 + \sum_{k} A_k sin(2\pi f_kt+\phi_k)
\end{equation}
where $A_0$ is the mean magnitude of the LC, and $f_k$, $A_k$, $\phi_k$ are the k-th frequency, its amplitude and phase, respectively. Only the frequency peaks with $A_k/\sigma_k > 4.5$ were accepted. The fitted series was subtracted from the LCs for prewhitening, and the residual was searched for the next dominating frequencies. The process was repeated until the noise remained.
P04 and our code gave similar results but we will present P04. 

There was a large gap of about a decade between SDSS and ZTF observations which produced frequency peaks at around 4000 d for a few BRRc stars. Considering their origin, we excluded these peaks from the results. 
In addition to the combined LCs, both SDSS and ZTF LCs were analysed separately, i.e. all 104 stars were analysed three times. 

\begin{figure*}
    \centering
    \includegraphics[width=15cm]{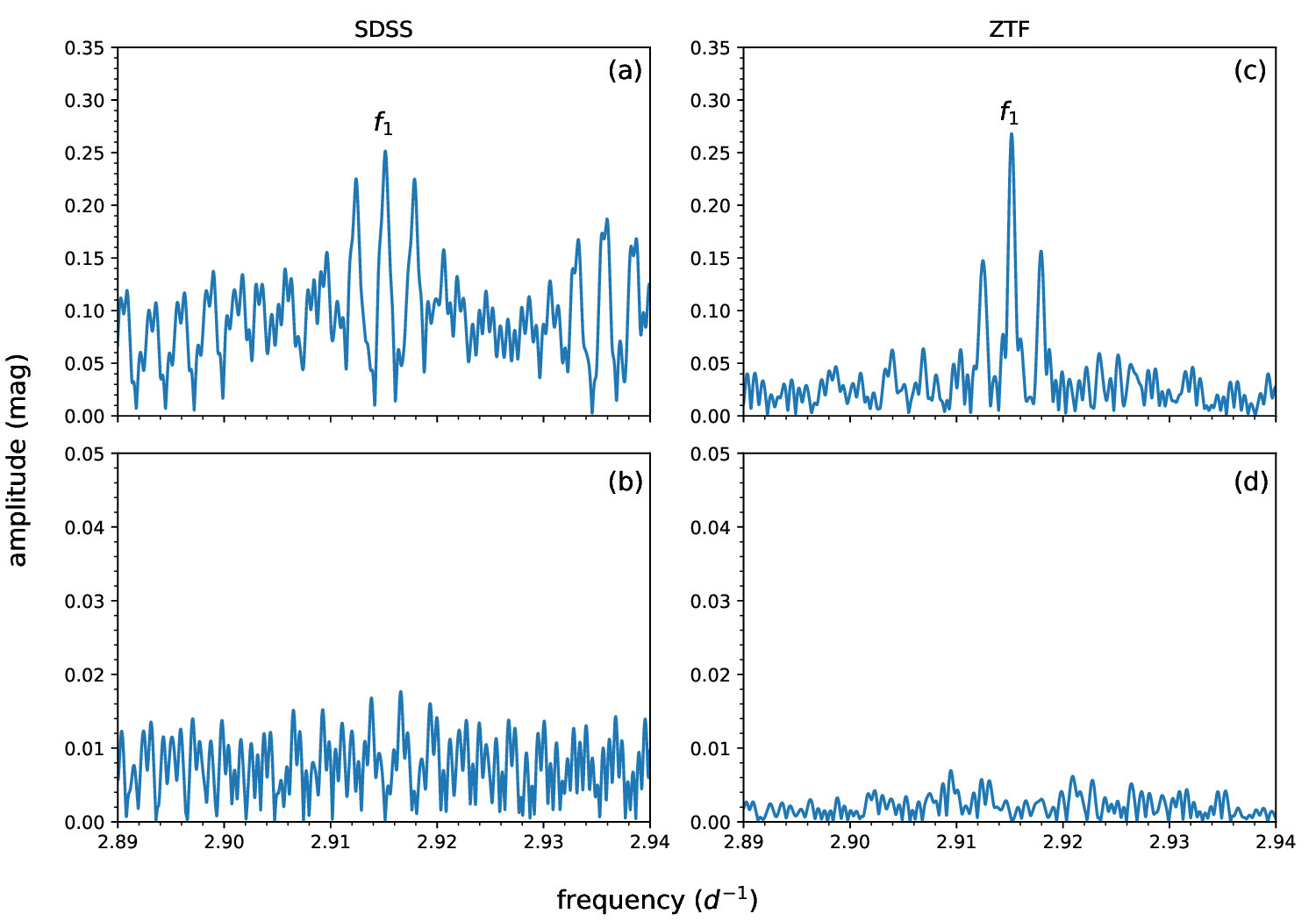}
    \caption{BRRc star SDSS ID 3585856: Fourier analysis of individual LCs from SDSS (left panels) and ZTF (right panels). In the left panels, (a) shows $f_1$ detected from SDSS LC, and (b) shows no Blazhko peaks but just the background noise. Similarly in the right panels, (c) shows $f_1$ detected from ZTF LC, and (d) shows no Blazhko peaks but just the background noise.}
    \label{fig:1}
\end{figure*}

The first step of the analysis in P04 was to determine the dominating pulsation frequency $f_1$. 
For many BRRc stars the expected pulsation frequency peak was above the Nyquist frequency which was in the range: 0.158 - 2.729 d$^{-1}$, except SDSS 376465 which had Nquist frequency of 218.208 d$^{-1}$. 
In many cases we had no choice but to search for frequencies above the Nyquist frequency; we can do this with due caution \citep{murphy2013super}. In this work, the frequencies were included if their single-to-noise ratio (SNR) $> 4.5$ as computed by P04. The main frequency $f_1$ and its significant number of (generally three) harmonics were subtracted from the LCs to prewhiten them. The second step was to search for the other significant frequency peaks in the prewhitened LCs. Each time, prewhitening included all the previously detected peaks subtracted simultaneously, to reduce the error.

\begin{figure}
    \centering
    \includegraphics[width=8cm]{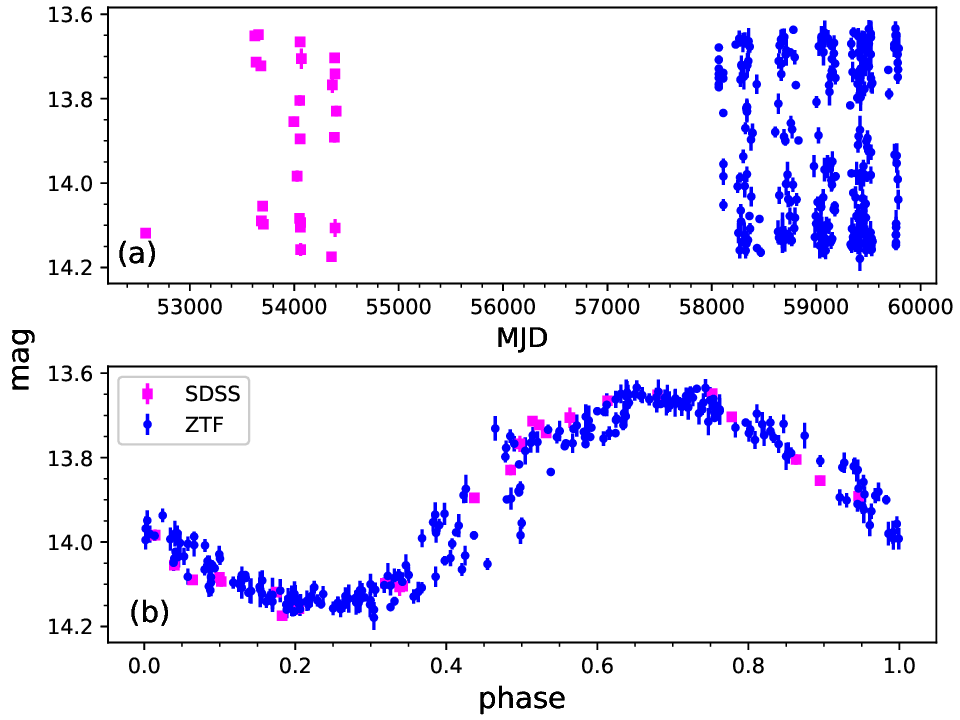}
    \includegraphics[width=8cm]{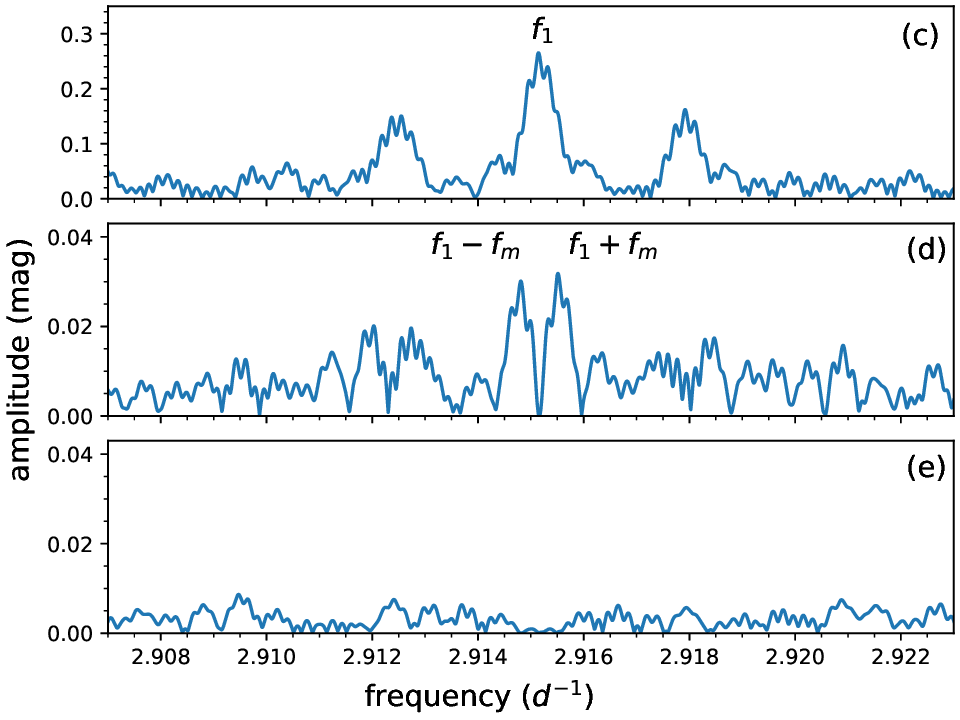}
    \caption{SDSS 3585856: Panels (a) and (b) show the raw data and folded LC, respectively. Magenta points are data from SDSS and blue points are from ZTF. Panels (c), (d), and (e) are the steps of prewhitening the LC shown as the Fourier spectrum. (c) shows the raw spectrum around the main frequency $f_1$, (d) shows the position of the Blazhko peaks $f_1\pm f_m$ in the residual spectrum after we prewhitened with the main frequency and its harmonics, and (e) is the noise after prewhitening the LC with all the previous frequencies.}
    \label{fig:8}
\end{figure}

When a Blazhko RRc LC is analysed using Fourier decomposition, the different Blazhko modulations appear as close frequency component(s) in the vicinity of radial mode frequency and its harmonics $kf_1$ ($k=1,2,\dots$) as $kf_1-f_{\mathrm m}$ and $kf_1+f_{\mathrm m}$. The modulation frequency $f_{\mathrm m}$ is called the Blazhko frequency, and its inverse is the Blazhko period $P_{\mathrm B}$. 
Thus longer the $P_B$, shorter the $f_{\mathrm m}$, which implies that the LC needs to be extended over a long time span, and has enough data points to detect long modulations. The side peaks, $kf_1-f_{\mathrm m}$ and $kf_1+f_{\mathrm m}$, can be equidistant or non-equidistant from $kf_1$, and can have asymmetric amplitudes \citep{benko2011, vanderplas2018}. For some Blazhko RRLs, there may be only one of the side peaks, while for some, multiple Blazhko modulations can be observed \citep{jurcsik2008extensive, chadid2010first}. 
The amplitudes of Blazhko peaks are around a few tenths of magnitudes, much smaller than that of $f_1$, and are difficult to detect in the presence of $f_1$. After prewhitening the LCs with $kf_1$, these Blazhko side peaks emerge in the Fourier spectrum. In the current work, if both the side peaks were detected, an average was taken as the final value of $f_{\mathrm m}$. If only one of the side peaks was observed, then that value was used to calculate $f_{\mathrm m}$. 

When the individual LCs from SDSS and ZTF were analysed, they did not always yield $P_{\mathrm B}$, or, if they did, the corresponding error was be high. One such example is seen in Fig~\ref{fig:1}. The left panels are Fourier spectra originated from SDSS data, while the right panels are that from ZTF data. No Blazhko peaks were detected in either of them but from the combined LC, we could detect the Blazhko peaks (Fig.~\ref{fig:8}). 

\begin{figure}
    \centering
    \includegraphics[width=8cm]{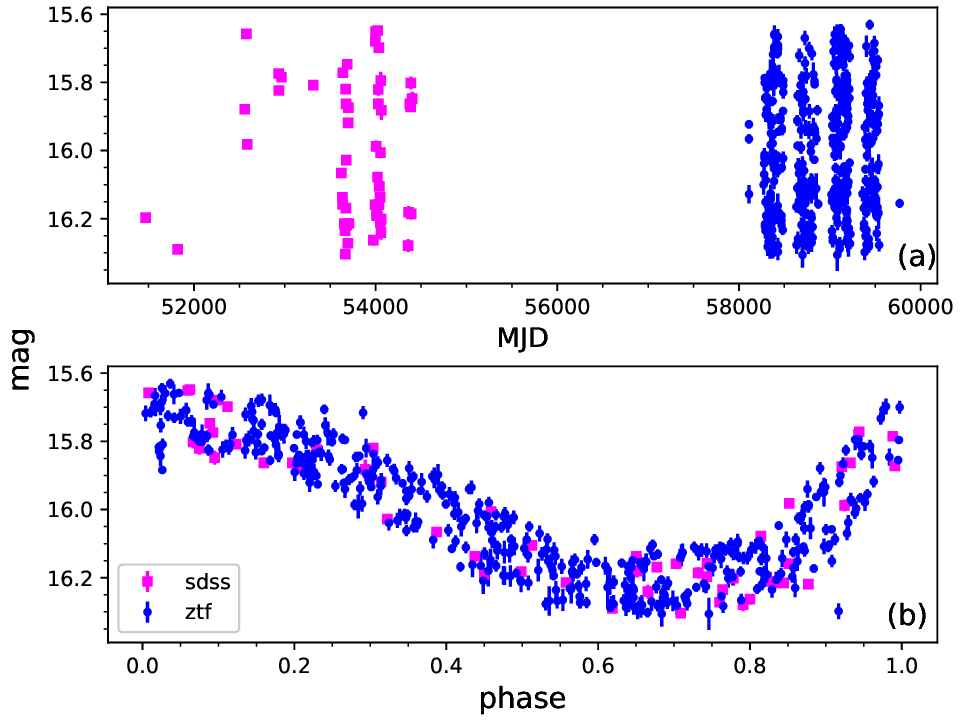}
    \includegraphics[width=8cm]{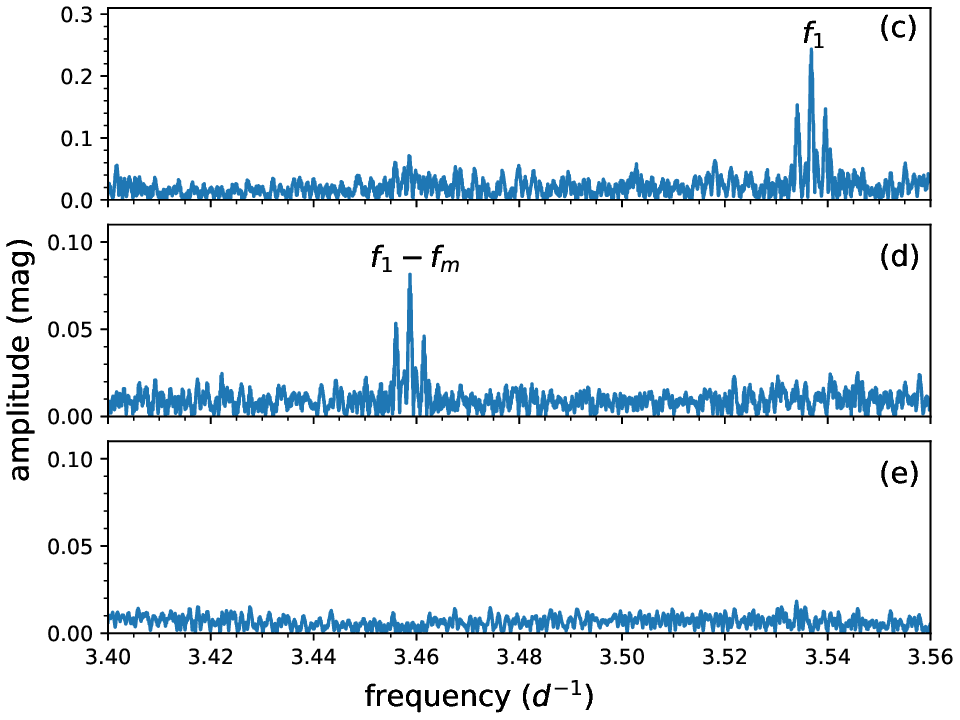}
   \caption{SDSS 747380: Panels (a) and (b) show the raw data and folded LC, respectively. Magenta points are data from SDSS, and blue ones are from ZTF. Panels (c), (d), and (e) are the steps of prewhitening. (c) shows the main spectrum around the main pulsation frequency $f_1$, (d) shows left hand side Blazhko peak $f_1-f_{\mathrm m}$ after we prewhitened the LC with $f_1$ and its significant harmonics, and (e) is the noise at the end of the prewhitening process.}
    \label{fig:7}
\end{figure}

\begin{figure}
    \centering
    \includegraphics[width=8cm]{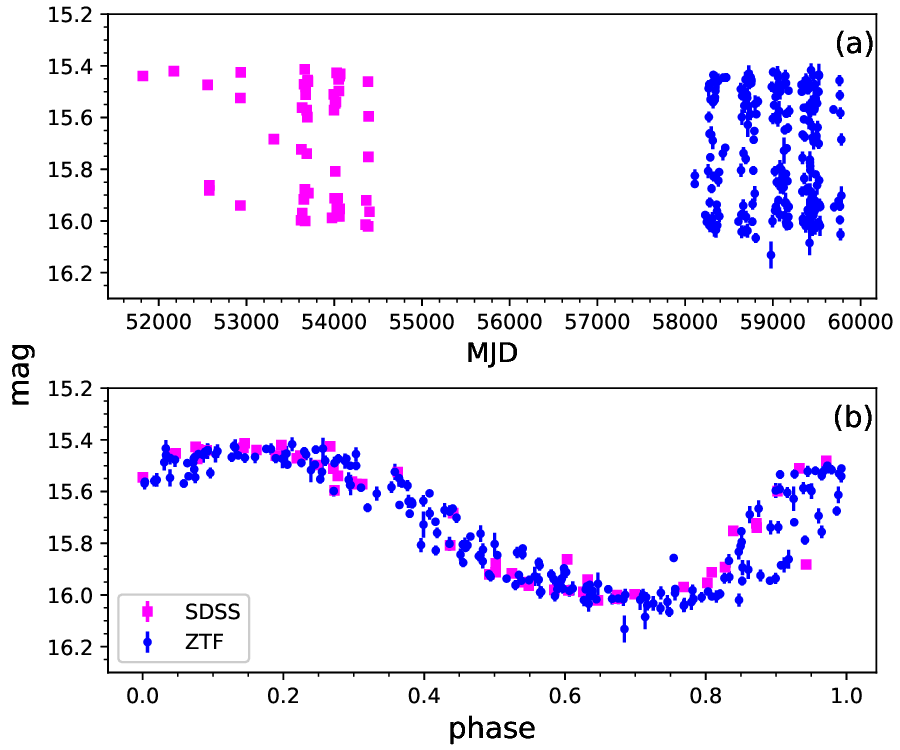}
    \includegraphics[width=8cm]{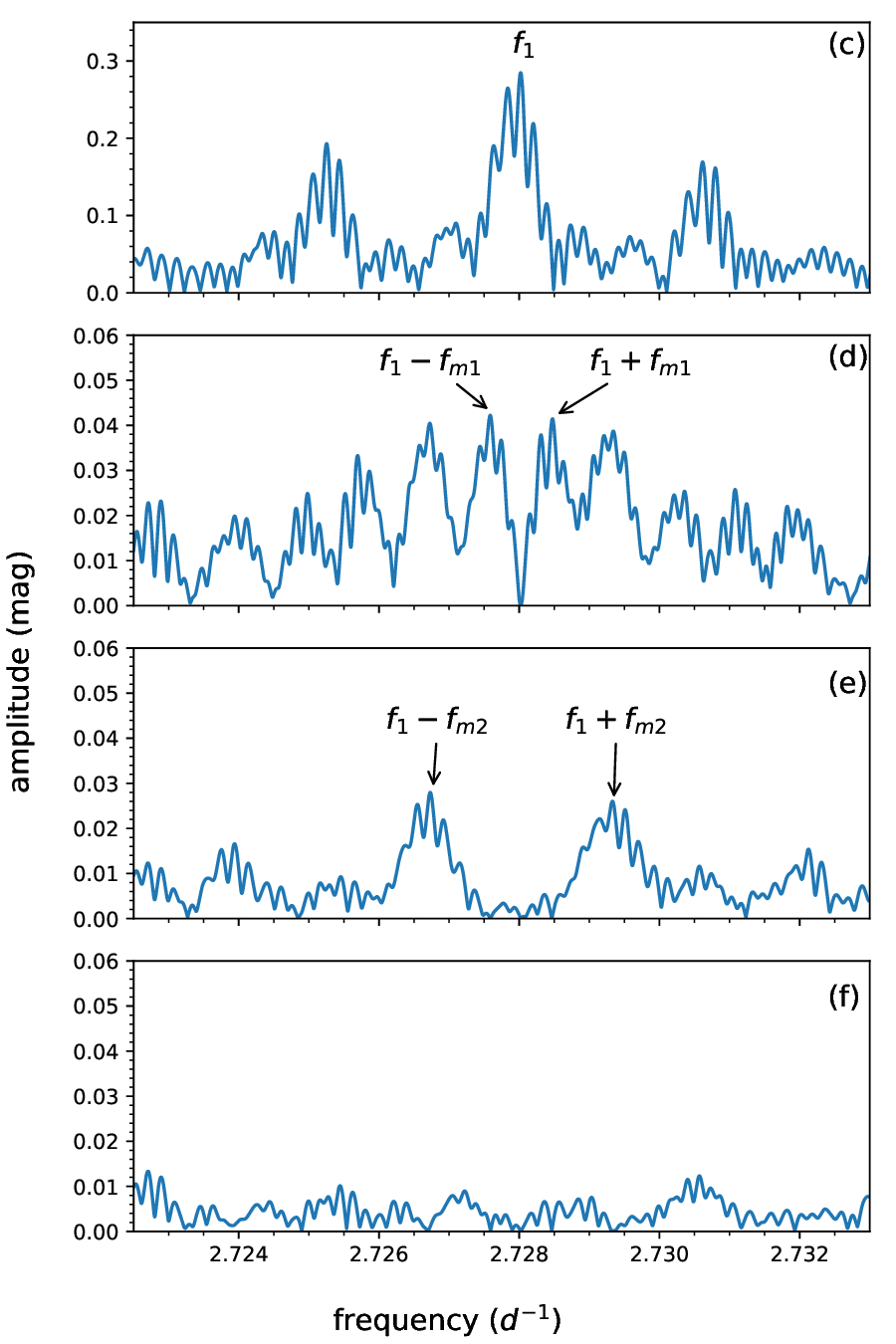}
    \caption{SDSS 2863787: panels (a) and (b) are similar to that of the Fig~\ref{fig:7}. (c) shows the Fourier spectrum around the main frequency, (d) shows the first pair of the Blazhko side peaks after we prewhitened the LC with $f_1$ and its significant harmonics, (e) shows the second pair after we prewhitened the LC with the first Blazhko pair, and (f) shows the noise at the end of the prewhitening process.}
    \label{fig:9}
\end{figure}

\begin{figure}
    \centering
    \includegraphics[width=8cm]{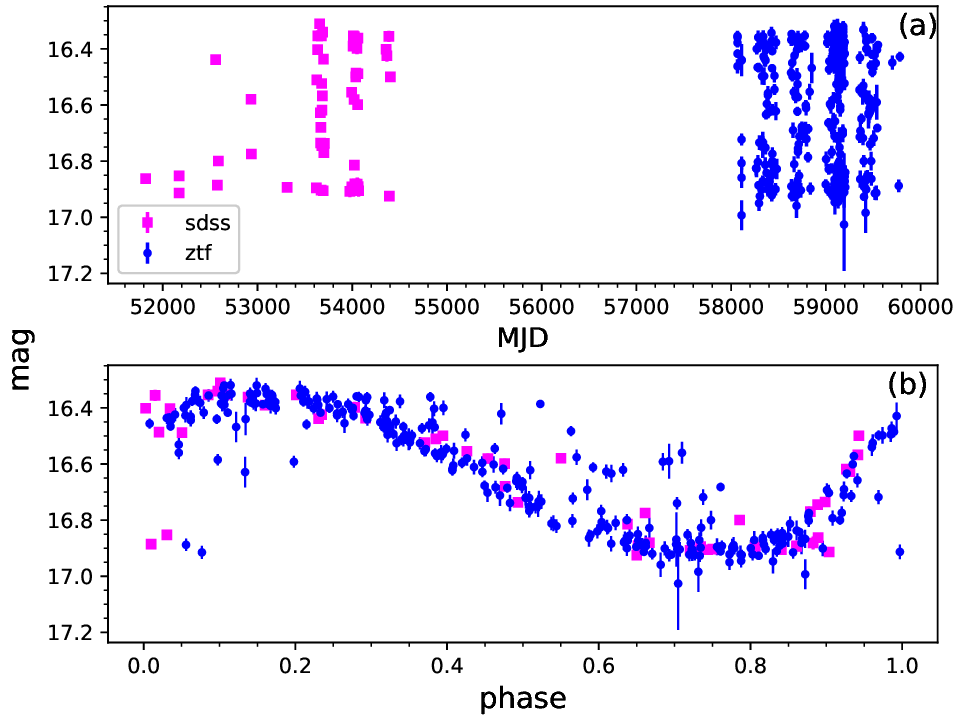}
    \includegraphics[width=8cm]{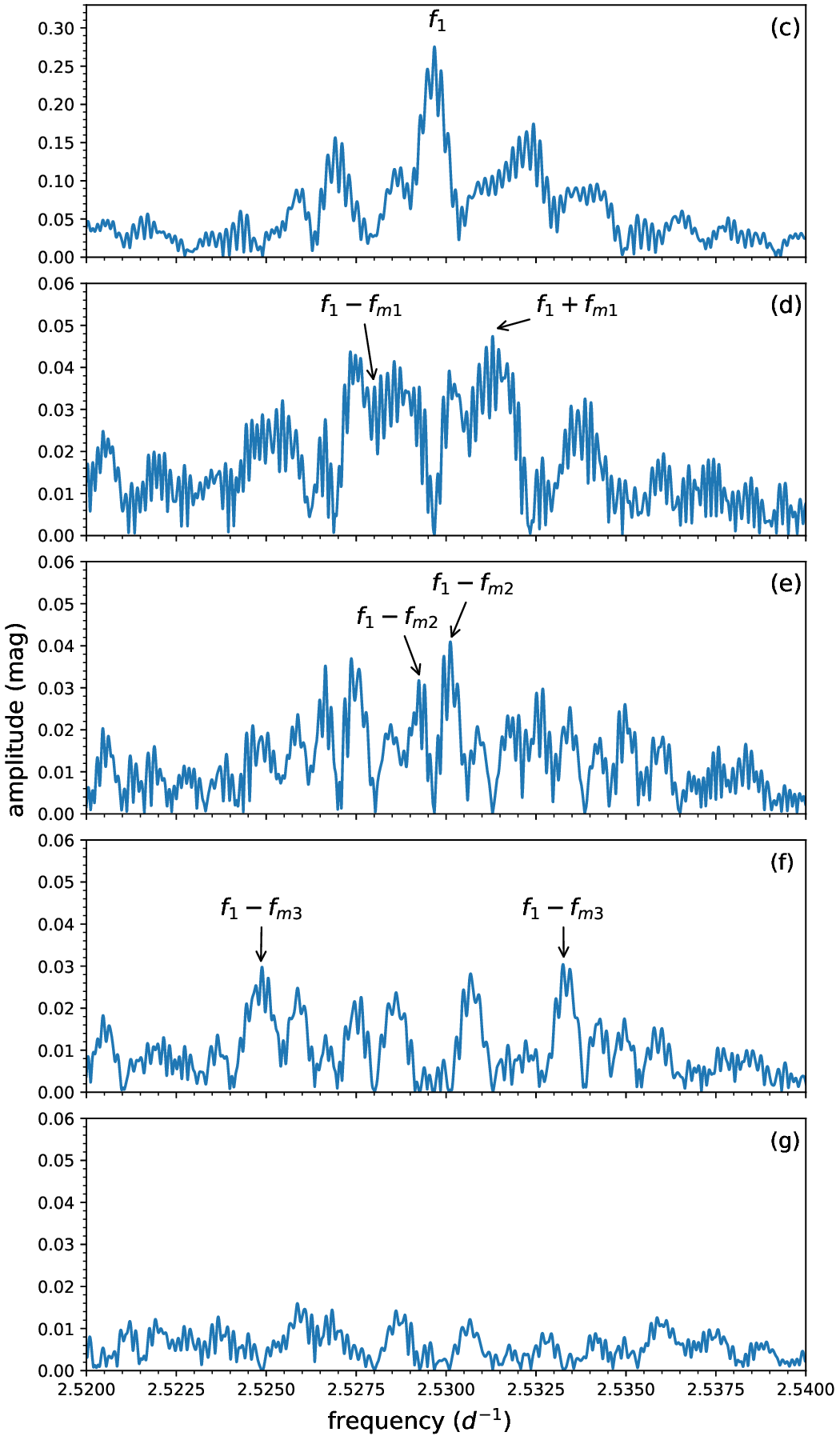}
    \caption{SDSS 1482164: The meaning of panels a)-e) are the same as in Fig.~\ref{fig:9}, panel f) shows the result of one more prewhitening step due to the third Blazhko modulation. Panel g) shows the end of the prewhitening.} 
    \label{fig:10}
\end{figure}

The steps of the analysis in different cases are illustrated in Figs.~\ref{fig:8}-\ref{fig:10}.
The top panels show the combined SDSS and ZTF data series. The second panels of the figures contain the LCs folded with the main period. The subsequent panels represent the Fourier spectra around the main frequencies from top to bottom, starting with the spectra of the original data series, followed by the spectra of the residual LCs prewhitened by the frequencies of one (Figs.~\ref{fig:8}-\ref{fig:7}), two (Fig.~\ref{fig:9}), and three modulations (Fig.~\ref{fig:10}). In all cases, the bottom panels show the noise after pre-whitening with all significant pulsation and modulation frequencies.

The error associated to each frequency was computed using Monte Carlo simulation in P04. Frequency error is in the range $9.5\times10^{-7}$ to $1.00\times 10^{-5}$~d$^{-1}$. Standard errors were computed for each $f_{\mathrm m}$ originated from the side peak(s), and were propagated when taking the average for $P_{\mathrm B}$.


\section{Results and Discussion}

For the 104 RRc stars published by S10 in the Stripe 82 region, our analysis showed that eight of them were rather RRd type. These stars are listed in Table~\ref{table:table1}. Their detailed analysis will be presented in a separate paper.

Blazhko side peaks were found for 34 of the remaining 96 stars of S10.
The main results on these stars are summarised 
in Table~\ref{table:table2}. It lists information on the newly identified BRRc stars including the main pulsation frequency $f_1$, its Fourier amplitude $A_1$ and phase $\phi_1$; the Blazhko period $P_{\mathrm B}$ and its $1\sigma$ error $\sigma(P_{\mathrm B})$.
The frequencies, amplitudes, and phases of the right and left hand side Blazhko peaks are also given.
The period $P$ of BRRc stars were found to be in the range of 0.261 to 0.432 d which sampled the instability strip rather well. 
For five stars, $P$ values published in S10 were incorrect because their data were suffering the one day alias problem. Using our extended LCs, the correct periods were determined, which can be confirmed via folding the LCs. These stars are marked with an asterisk in Table~\ref{table:table2}. 

\begin{table}
\centering
\begin{tabular}{llll}
\hline
SDSS ID & $f_0$ & $f_1$ & ratio \\
    & (d$^{-1}$) & (d$^{-1}$) \\
\hline
1078860 & 1.886082 & 2.528637 & 0.746 \\
3478713 & 2.047186 & 2.746912 & 0.745  \\
850835 & 1.826386 & 2.449997 & 0.745  \\
3214909 & 1.977810 & 2.655536 & 0.745  \\
2488976 & 1.842614 & 2.468807 & 0.746  \\
2291937 & 1.923916 & 2.577745 & 0.746  \\
2464128 & 1.878133 & 2.514649 & 0.747  \\
2506078 & 2.060011 & 2.769921 & 0.744 \\ 
\hline
\end{tabular}
\caption{RRd stars discovered in the Stripe 82 region which were misclassified by S10 as RRc stars. The columns are: SDSS ID, frequency of the fundamental mode $f_0$, the first overtone $f_1$, and their ratio.}
\label{table:table1}
\end{table}

\begin{figure}
    \centering
    \includegraphics[width=9.5cm]{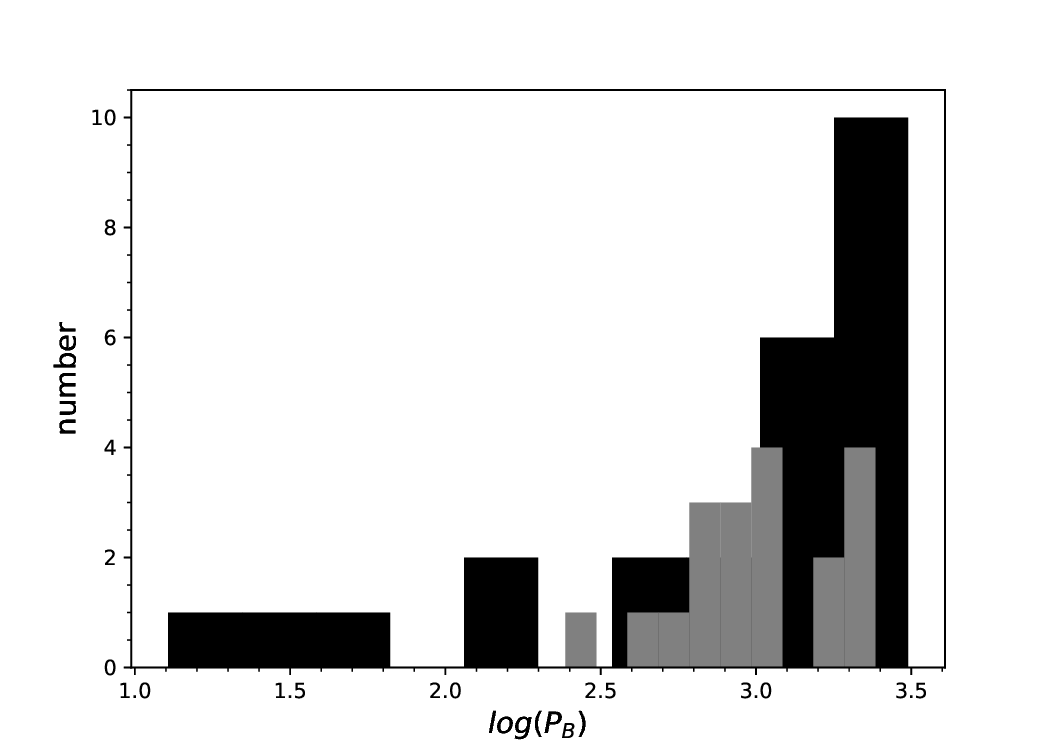}
    \caption{Histogram of the Blazhko period $P_{\mathrm B}$. Black and gray colors represent BRRc stars with single and multiple $P_{\mathrm B}$, respectively.}
    \label{fig:3}
\end{figure}

The $P_{\mathrm B}$ is in the range of 12.808 and 3088~d. The latter value (although with a large error) is the longest Blazhko period ever detected for an RRc star. The previous record holder was OGLE RRLYR-02478 with a period of 2954~d \citep{netzel2018}. 
Out of 34 BRRc stars, 25 (74\%) show single, and 9 (26\%) show multiple Blazhko modulations. The distribution of $P_{\mathrm B}$ is shown in Fig~\ref{fig:3}, where black and gray colors correspond to BRRc stars with single and multiple Blazhko effect, respectively. For stars with multiple Blazhko effect, $P_{\mathrm B}$ was greater than 200~d. We did not find any BRRc stars with multiple short Blazhko periods.

Our 34 stars showing the Blazhko effect represent a 35.42\% incidence rate.
It is higher than any previously documented rates. \citet{netzel2018} compiled the incidence rates of the Blazhko effect in RRc stars for different stellar systems, and found it between 4 and 19\% (see their table 4). Recently, \cite{benko2023time} added to the list the missing rate of the Galactic field as 10.4\%.
It is noteworthy that the higher incidence rates belong to metal-poor globular clusters (M3: 12\%, \citealt{jurcsik2014}; NGC 6362: 19\%, \citealt{smolec2017}). On the other hand, \citet{Smolec2005} showed that the different Blazhko incidence rates of different stellar systems are not a simple function of metallicity. Still, for the Galactic halo with the metallicity similar to that of the globular clusters, a higher incidence rate was to be expected but such a high value was surprising.
However, 46\% of all BRRc stars with single or multiple $P_{\mathrm B}$, show very long period Blazhko effect ($P_{\mathrm B} > 1000$~d) which can be also seen in Fig~\ref{fig:3}. This fact might be the key to why we obtained a much higher incidence rate from out long data series than in previous works.

\begin{figure}
    \centering
    \includegraphics[width=9.5cm]{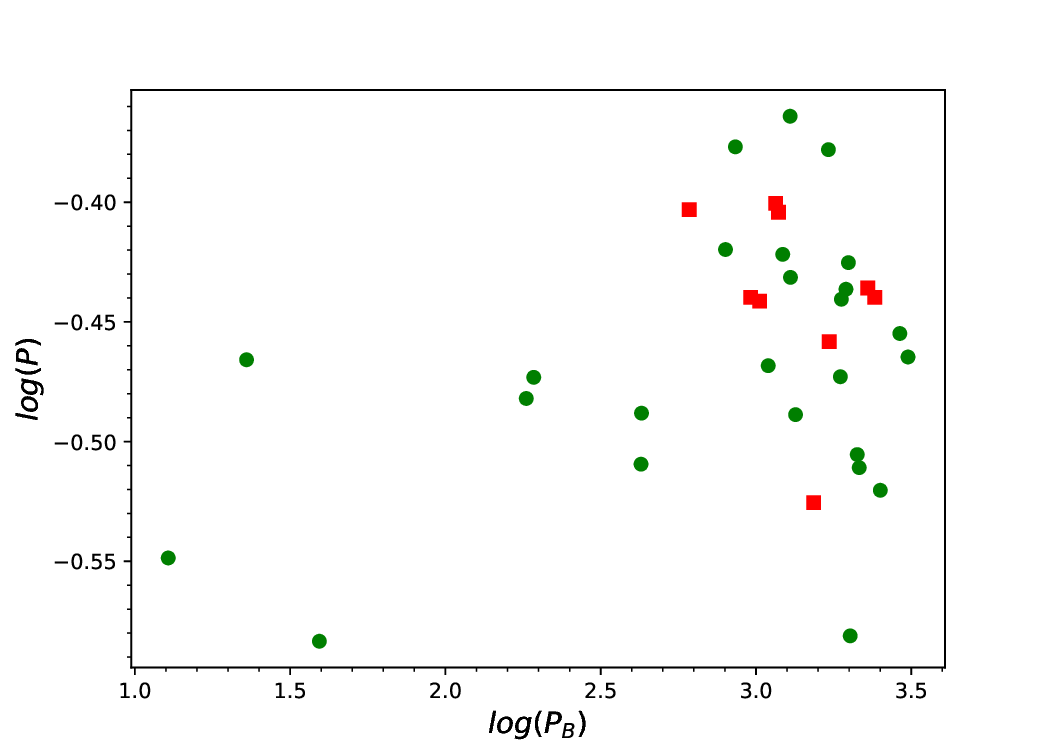}
    \caption{Plot of the main pulsation period P to the Blazhko period $P_{\mathrm B}$ in log-log scale. Green and red points show the BRRc stars with single and multiple Blazhko effect, respectively.}
    \label{fig:4}
\end{figure}

We plotted $P$ against $P_B$ in Fig~\ref{fig:4} to check the presence of any correlation. \cite{jurcsik2005distribution} found a tendency between the two  quantities, such that RRLs with periods $P < 0.4$~d had comparatively shorter $P_{\mathrm B}$ of order of some days. 
\cite{szczygiel2007multiperiodic} detected a bimodal distribution for the modulation period of the RRc stars in the ASAS Survey. 
The two maxima of the distribution are around $\sim10$ and $\sim1500$ days, and there is a gap between the two groups (between $\sim$20 and $\sim$300 d) where there are hardly any stars. 
It has been suggested that there may be two different physical origins behind the variations in the two groups. If it is true and the stars with `classical' Blazhko effect are only in the group of 10 days, and the long period group is something else, then results of \cite{szczygiel2007multiperiodic} do not necessarily contradict the finding of \citet{jurcsik2005distribution}.
A slight bimodality can also be recognized in the combined OGLE I-IV data, but due to the significantly long data series available for only a few stars, the short part is much stronger \citep[see Fig. 6 in][]{netzel2018}.
Our data series are mainly sensitive to the long periods, so in Fig.~\ref{fig:4} we have mostly obtained the long-period part of the bimodal distribution.
\begin{table*}
\hskip-2.5cm
\begin{tabular}{llllllllllll}
\hline
SDSS ID & $f_1$ & $P_{\mathrm B}$ & $\sigma(P_{\mathrm B})$ & $A_1$ & $\phi_1$ & $f_+$ & 
$A_+$ & $\phi_+$ & $f_-$ & $A_-$ & {$\phi_-$} \\ 
        & (d$^{-1}$) & (d) & (d) & (mag) & (deg) & (d$^{-1}$) & (mag) & (deg) & (d$^{-1}$)  & (mag) & (deg) \\
\hline
747380*  & 3.536821 & 12.808  & 0.001   & 0.251    & 0.616  & -        & -        & -      & 3.458743 & 0.084    & 0.706  \\
562035   & 2.922886 & 22.889  & 0.003   & 0.230    & 0.500  & 2.966577 & 0.070    & 0.860  & 2.879197 & 0.122    & 0.668  \\
4614863* & 3.831802 & 39.28   & 0.03    & 0.249    & 0.459  & 3.857263 & 0.050    & 0.244  & 3.806343 & 0.055    & 0.103  \\
2659801  & 3.033801 & 182     & 1       & 0.320    & 0.836  & 3.03928  & 0.016    & 0.393  & 3.028298 & 0.018    & 0.147  \\
638540   & 2.972634 & 192     & 1       & 0.245    & 0.138  & 2.977831 & 0.014    & 0.896  & 2.967441 & 0.008    & 0.634  \\
2609442  & 3.231722 & 426     & 3       & 0.269    & 0.427  & 3.234067 & 0.013    & 0.939  & -        & -        & -      \\
3681504  & 3.076957 & 428     & 4       & 0.279    & 0.756  & 3.079215 & 0.016    & 0.832  & 3.074537 & 0.022    & 0.701  \\
288008*  & 2.628844 & 798     & 10      & 0.194    & 0.734  & 2.630027 & 0.113    & 0.596  & 2.627511 & 0.065    & 0.270  \\
3348773  & 2.381611 & 858     & 15      & 0.238    & 0.839  & 2.382785 & 0.038    & 0.174  & 2.380455 & 0.037    & 0.853  \\
1217801  & 2.939625 & 1095    & 32      & 0.243    & 0.536  & 2.940526 & 0.012    & 0.226  & 2.9387   & 0.015    & 0.790  \\
959802   & 2.641075 & 1219    & 22      & 0.247    & 0.885  & 2.642081 & 0.042    & 0.110  & 2.640383 & 0.045    & 0.746  \\
376465   & 2.312534 & 1288    & 26      & 0.239    & 0.961  & 2.313314 & 0.016    & 0.239  & 2.311762 & 0.015    & 0.733  \\
4383295  & 2.700236 & 1292    & 19      & 0.266    & 0.489  & 2.701011 & 0.050    & 0.386  & 2.699463 & 0.047    & 0.939  \\
2314306  & 3.081313 & 1341    & 18      & 0.272    & 0.509  & 3.082061 & 0.106    & 0.146  & 3.08057  & 0.105    & 0.083  \\
3358190  & 2.388038 & 1711    & 29      & 0.229    & 0.303  & 2.388621 & 0.052    & 0.915  & 2.387453 & 0.034    & 0.327  \\
1478867  & 2.970986 & 1870    & 77      & 0.257    & 0.748  & 2.971514 & 0.028    & 0.191  & 2.970444 & 0.017    & 0.696  \\
3397977  & 2.757541 & 1884    & 83      & 0.223    & 0.019  & 2.758071 & 0.028    & 0.875  & 2.757009 & 0.047    & 0.711  \\
3031571  & 2.731217 & 1949    & 42      & 0.306    & 0.753  & 2.731697 & 0.040    & 0.967  & 2.730667 & 0.041    & 0.144  \\
3308790  & 2.662176 & 1985    & 38      & 0.225    & 0.041  & 2.66268  & 0.078    & 0.132  & 2.661672 & 0.066    & 0.427  \\
765345   & 3.811996 & 2011    & 87      & 0.153    & 0.930  & 3.812496 & 0.015    & 0.392  & 3.811501 & 0.022    & 0.659  \\
1091627  & 3.201902 & 2119    & 41      & 0.250    & 0.220  & 3.202372 & 0.051    & 0.614  & 3.201428 & 0.053    & 0.624  \\
1986301  & 3.242536 & 2150    & 35      & 0.298    & 0.716  & 3.242995 & 0.082    & 0.175  & 3.242065 & 0.093    & 0.567  \\
2213142  & 3.313901 & 2514    & 112     & 0.248    & 0.505  & 3.314263 & 0.028    & 0.268  & 3.313459 & 0.024    & 0.081  \\
429508   & 2.85003  & 2906    & 148     & 0.245    & 0.523  & 2.850376 & 0.059    & 0.151  & 2.849688 & 0.034    & 0.113  \\
3585856  & 2.915149 & 3088    & 126     & 0.244    & 0.039  & 2.915485 & 0.031    & 0.203  & 2.914836 & 0.031    & 0.008  \\
4455741  & 2.752802 & 962     & 8       & 0.215    & 0.724  & -        & -        & -      & 2.751762 & 0.041    & 0.590  \\
         &          & 2412    & 39      &          &        & 2.753207 & 0.099    & 0.011  & 2.752377 & 0.071    & 0.026  \\
2396176* & 2.762629 & 1027    & 4       & 0.179    & 0.632  & -        & -        & -      & 2.761655 & 0.128    & 0.486  \\
         &          & 431     & 3       &          &        & 2.765078 & 0.048    & 0.306  & 2.760429 & 0.062    & 0.643  \\
2639854  & 3.353221 & 1534    & 23      & 0.138    & 0.314  & 3.353902 & 0.062    & 0.944  & 3.352596 & 0.071    & 0.856  \\
         &          & 910     & 13      &          &        & 3.354286 & 0.035    & 0.926  & 3.352084 & 0.048    & 0.648  \\
2445511  & 2.536248 & 1182    & 18      & 0.259    & 0.438  & 2.537088 & 0.095    & 0.152  & 2.535396 & 0.059    & 0.980  \\
         &          & 712     & 13      &          &        & 2.537599 & 0.046    & 0.642  & 2.534785 & 0.060    & 0.226  \\
2175525  & 2.514798 & 1158    & 11      & 0.220    & 0.676  & 2.515677 & 0.062    & 0.832  & 2.513949 & 0.083    & 0.286  \\
         &          & 706     & 12      &          &        & 2.516243 & 0.030    & 0.722  & 2.513409 & 0.037    & 0.796  \\
4455741* & 2.752802 & 2412    & 39      & 0.215    & 0.724  & 2.753207 & 0.099    & 0.011  & 2.752377 & 0.071    & 0.026  \\
         &          & 962     & 8       &          &        & -        & -        & -      & 2.751762 & 0.041    & 0.590  \\
1747387  & 2.872223 & 1722    & 13      & 0.140    & 0.519  & 2.872804 & 0.069    & 0.328  & -        & -        & -      \\
         &          & 1124    & 22      &          &        & 2.873118 & 0.039    & 0.257  & 2.871339 & 0.020    & 0.952  \\
2863787  & 2.728025 & 2291    & 71      & 0.289    & 0.571  & 2.728464 & 0.036    & 0.119  & 2.72759  & 0.044    & 0.287  \\
         &          & 750     & 14      &          &        & 2.729367 & 0.022    & 0.658  & 2.726701 & 0.026    & 0.988  \\
1482164  & 2.529674 & 610     & 9       & 0.276    & 0.472  & 2.531294 & 0.031    & 0.649  & 2.528014 & 0.031    & 0.864  \\
         &          & 2426    & 46      &          &        & 2.530091 & 0.119    & 0.274  & 2.529266 & 0.077    & 0.688  \\
         &          & 244     & 1       &          &        & 2.53327  & 0.045    & 0.075  & 2.524887 & 0.038    & 0.995  \\ \hline
\end{tabular}
\caption{Blazhko RRc stars discovered in the Stripe 82 region. The columns are: SDSS ID; $f_1$ is the main frequency; $P_{\mathrm B}$ is the Blazhko period; $\sigma(P_{\mathrm B})$ is the error associated with $P_{\mathrm B}$; $A_1$ and $\phi_1$ are the amplitude and phase of $f_1$; $f_+$ is the right hand side peak ($f_+=f_1+f_m$); $A_+$ and $\phi_+$ are the amplitude and phase of $f_+$; $f_-$ is the left hand side peak  ($f_-=f_1-f_m$); $A_-$ and $\phi_-$ are the amplitude and phase of $f_-$. If any side peak is not detected, '-' is written instead. * after the SDSS ID indicates stars whose main periods were one day alias in S10.}
\label{table:table2}
\end{table*}

\begin{figure}
    \centering
    \includegraphics[width=9cm]{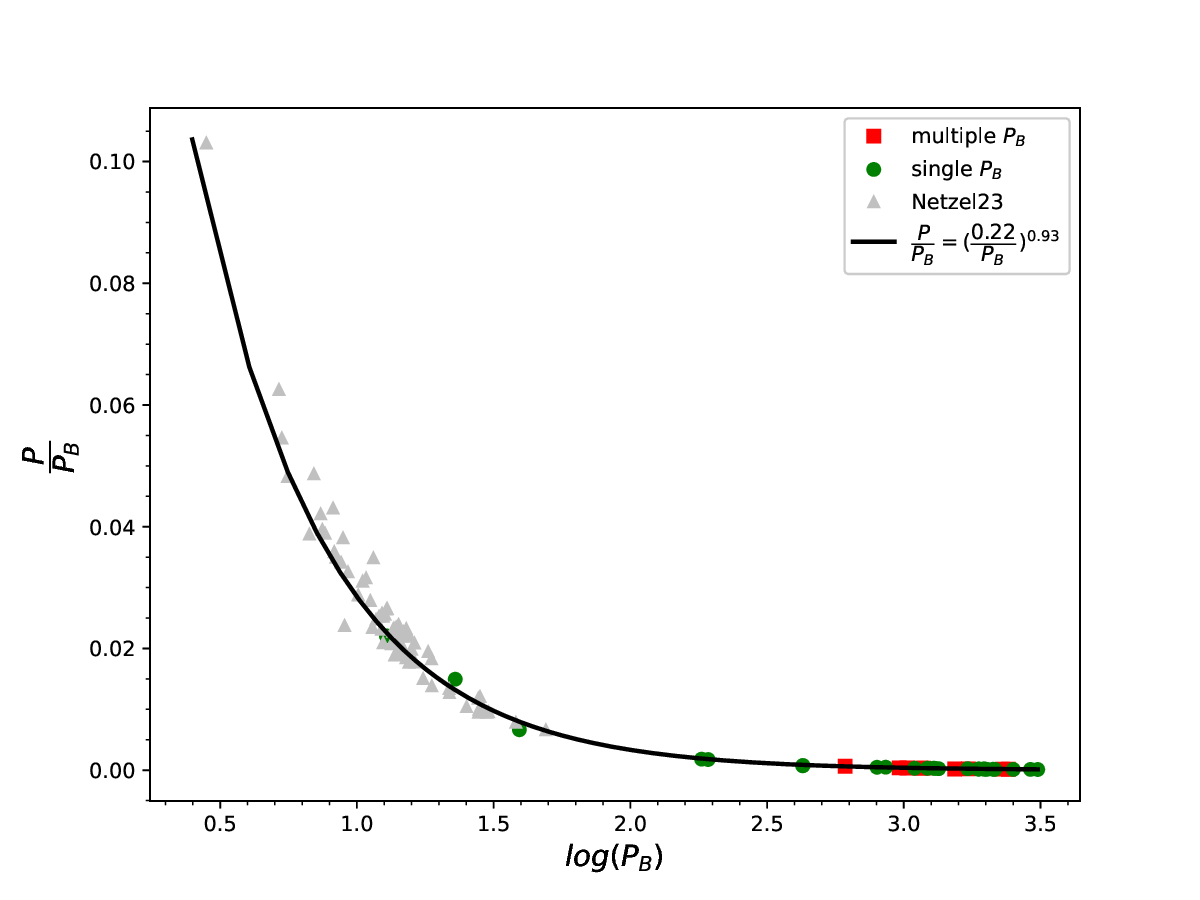}
    \caption{Ratio of $P$ to $P_{\mathrm B}$ plotted against $\log(P_{\mathrm B})$. The green and red points represent BRRc stars with single and multiple $P_{\mathrm B}$, respectively, and the silver points are \cite{Netzel2023} data. The black curve shows the fitting.}
    \label{fig:5}
\end{figure}

What could be the reason for the bimodality for RRc stars (there is no such evidence for RRab stars)? 
From one year of continuous measurements of the TESS, \citet{benko2023time} concluded that RRc stars in which the $f_{61}$ and $f_{63}$ non-radial mode frequencies (see their definition in the cited work) appear simultaneously, typically showed long-period semi-regular phase variations. These phase variations are indistinguishable from the long-period Blazhko variations found in the present work, or even in the ASAS or OGLE surveys. If this hypothesis is correct, then the Blazhko incidence rate found would actually be $\sim$5\%, and the number of RRc stars where the two frequencies of the non-radial mode are excited simultaneously, would be $\sim$30\%. The difference in the distributions of RRab and RRc stars would naturally be explained by the fact that we have never found frequencies similar to $f_{61}$ and $f_{63}$ in RRab stars. This hypothesis can be tested in the future if TESS provides data series of sufficient length (10\textendash15 years) and quality to detect additional frequencies.

Fig~\ref{fig:5} shows the ratio of $P$ to $P_{\mathrm B}$ plotted against $P_{\mathrm B}$ in log-scale. The green points are BRRc stars with single $P_{\mathrm B}$, and the red points are those with multiple $P_{\mathrm B}$, both showing a decreasing trend which can be interpreted as $P$ being inversely proportional to $P_{\mathrm B}$. Using the curve fitting method, and keeping the dimensions in mind, we fit an empirical equation: 
\begin{equation}\label{eq:3}
    \frac{P}{P_B} = \left(\frac{k}{P_B}\right)^\epsilon
\end{equation}
where $k = 0.22\pm 0.024$~d, and $\epsilon = 0.93\pm 0.002$. We gave k the dimension of days, and kept $\epsilon$ dimensionless to satisfy the dimension equation. 
Fitting of this equation is shown as the black curve in Fig~\ref{fig:5}. We do not associate this equation with any physical phenomena occurring in RRc stars. \cite{netzel2018} (see their fig. 7) observed such ratios to be in a wide range of values but their scatter hid the decreasing trend observed here. We added the most recent data from \cite{Netzel2023} to the plot as the silver points, and conclude that the same trend was followed, with \cite{Netzel2023} data falling on the left side, while our data is on the right. This illustrates that we have indeed been able to capture the long period part of the bimodal distribution of \cite{szczygiel2007multiperiodic}.

\section{Summary}
S10 classified 104 RRLs as RRc stars. Our study shows 8 of them are RRd stars. 
In the remaining 96 RRc stars, 34 display Blazhko effect giving the incidence rate of 35.42\% which is higher than any previously published values. We briefly discussed the possible reasons for this.
The shortest Blazhko period ($P_{\mathrm B}$) found is $12.808\pm0.001$~d for SDSS 747380, 
while the longest is $3100\pm126$~d for SDSS 3585856. 
The latter is the longest $P_{\mathrm B}$ ever detected. The vast majority (85\%) of detected $P_{\mathrm B}$ are above 200 days. This illustrates well the potential of the extreme length of the combined SDSS and ZTF data we use.
The 8 BRRc stars show two Blazhko modulations, and one, SDSS 1482164, shows three modulations.  
Period ratio $P$ to $P_B$ shows a steep decreasing trend with increasing $P$ due to a large number of long $P_{\mathrm B}$, which indicates that $P$ is inversely proportional to $P_B$. 

\section{Acknowledgments}
The author thanks the reviewers for their comments.
We are thankful for funding from the National Science and Technology Council (Taiwan) under the contract 109-2112-M-008-014-MY3.
JMB was partially supported by the ‘SeismoLab’ KKP-137523 \'Elvonal and NN-129075 grants of the Hungarian Research, Development and Innovation Ofﬁce (NKFIH).
Based on observations obtained with the Samuel Oschin Telescope 48 inch Telescope at the Palomar Observatory as part of the Zwicky Transient Facility project. ZTF is supported by the National Science Foundation under grant No. AST1440341 and AST-2034437, and a collaboration including current partners Caltech, IPAC, the Weizmann Institute of Science, the Oskar Klein Center at Stockholm University, the University of Maryland, Deutsches Elektronen-Synchrotron and Humboldt University, the TANGO Consortium of Taiwan, the University of Wisconsin at Milwaukee, Trinity College Dublin, Lawrence Livermore National Laboratories, IN2P3, University of Warwick, Ruhr University Bochum, Northwestern University and former partners the University of Washington, Los Alamos National Laboratories, and Lawrence Berkeley National Laboratories. Operations are conducted by COO, IPAC, and UW.

\bibliography{ref.bbl}

\end{document}